# Impact of Imprecision of the Time Delay on Imaging Result in Confocal Algorithm

Wenyi Shao[1][2][3], Beibei Zhou*[2][3], and Gang Wang[2]

(1) University of Georgia, Athens, GA30602
(2) Jiangsu University, Zhenjiang, PR. China
(3) University of North Carolina at Charlotte, Charlotte, NC 28262
E-mail: wyshao@uga.edu

## Introduction

The confocal microwave imaging (CMI) algorithm, which is used in radar imaging, has been applied to microwave medical imaging (MMI) [1, 2]. In MMI, the diseased region to be imaged has different electromagnetic characteristics from that of the surroundings. Time -domain MMI system often radiates an ultra-wideband pulse and then collect the backscatter signals with an antenna array. The time delay is then compensated for every unit in the array according to the time flight to each focal point within the region of interest. The compensated signals are finally summed to calculate the pixel value of each focal point. This process is repeated to achieve the value for all focal points in a 3D space to reconstruct the image. Since the method is only constrained in the time domain, the complexity is minimal, and an image can be obtained usually in few seconds. This paper discusses the robustness of CMI in MMI. Specifically, we analyzed the variation of image contrast as errors occur in the time-delay calculation. A simplified model of breast cancer detection was implemented to study the relationship between time-shift errors and the impact on images, quantitatively.

## Simulation and Test

The model is a cuboid with 120 *mm* long and also 120 *mm* wide, shown in Fig. 1. The target is a spherical tumor ($\varepsilon_r$ = 50, σ = 4) with 4 *mm* diameter and is buried in medium (typical breast tissue, $\varepsilon_r$ = 9 and σ= 0.4). Its center locates at (60, 60, 50). Above the medium surface (z = 75) is air (no coupling medium [3] in use for the simplicity reason in this paper). A ultrawide band pulse with bandwidth 4.8 GHz, is transmitted by a transmitter which is 2 mm above the medium surface (the use of multiple transmitters has been explored in another paper [4]). And 25 receivers are located in the same plane (z = 77), forming a synthetic array [5]. The red bright point in the center of the image represents a tumor (multiple-tumor case has been explored in [6]). The simulation was processed by a self-developed software based on the finite difference time domain (FDTD) approach [7, 8]. The confocal algorithm representing the energy summed at $\vec{r}$ is calculated

$$E(\vec{r}) = [\sum_{m=1}^{M} T_m(t_m(\vec{r}))]^2 \qquad (1)$$

where $t_m(\vec{r})$ denotes time delay on the *m*th receiver and $T_m(t_m(\vec{r}))$ denotes the signal after a time shift or compensation. Therefore, there is a need of accurate evaluation of the propagation time from the target point to the detectors. Failure to do so may result in the degradation of image contrast. As such, the image quality depends on how well the signals are summed at $\vec{r}$, in other words, on the evaluation accuracy of the time delay. In our test, we adjusted the accuracy level of time-delay then linked to the image contrast. Some imaging results with increasing time-delay percentage error (TDPE) corresponding to the reduced contrast is given in Fig. 2. Ratio of signal to mean-noise ($S/\bar{N}$) is applied to represent the contrast. In this paper, we are only caring about contrast, because the impact on resolution has been explored in another paper [9].

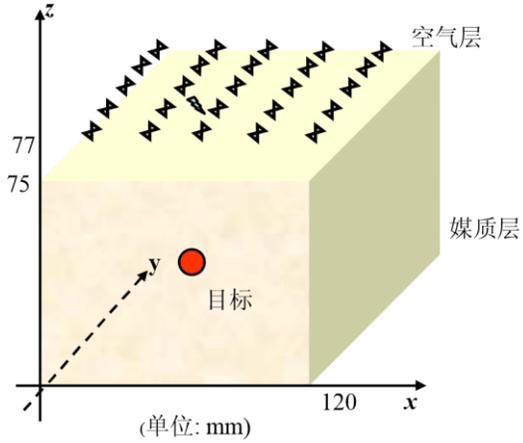
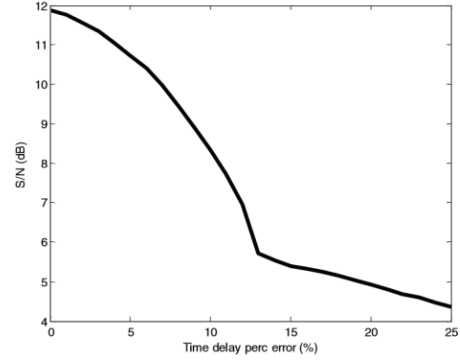

Fig. 3. Contrast varying with the time-delay percentage error.

Fig.1. The simplified model. ⚡ receiver ⚡ transmitter

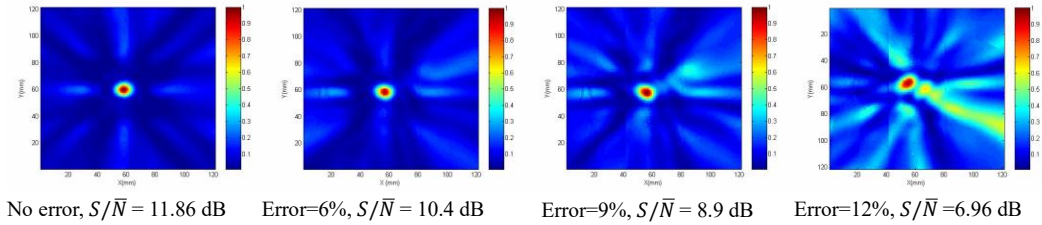

No error, $S/\bar{N}$ = 11.86 dB    Error=6%, $S/\bar{N}$ = 10.4 dB    Error=9%, $S/\bar{N}$ = 8.9 dB    Error=12%, $S/\bar{N}$ =6.96 dB

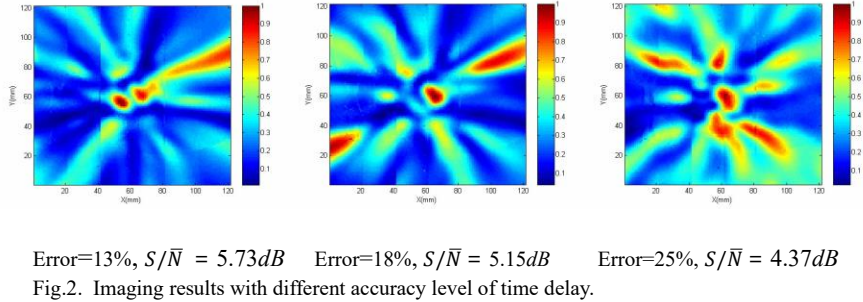

Error=13%, $S/\bar{N}$ = 5.73$dB$    Error=18%, $S/\bar{N}$ = 5.15$dB$    Error=25%, $S/\bar{N}$ = 4.37$dB$
Fig.2. Imaging results with different accuracy level of time delay.

Fig. 3 presents the relation of time-delay percentage error and contrast, denoted by $S/\bar{N}$ based on 20 groups of simulation data. It can be found that contrast decreases fast as the TDPE increases until it improves to 13%. Although the contrast goes down slowly after TDPE exceeds 13%, it is difficult to recognize the tumor's location from the image (Fig.2 error=18% and error=25%). This denotes the essential reason of contrast degradation: the existence of the time-delay error. The time delay equals the distance between target and the detector divided by the signal's travelling speed. Usually, the followings may cause the time-delay error:

(1) Antennas' location setting imprecisely. As the ground-penetrating detection be concerned, antenna (array) carried by vehicles or aircrafts has oscillations when it is moving and sampling data. This leads to the distance between antenna and imaging point cannot be evaluated precisely. Medical imaging has a even stricter requirement. Antennas have to be relocated quite often. Some mechanical precision problems cause the antenna unable to be relocated precisely which makes the time delay difficult to evaluate very correctly.

(2) Velocity of detecting signal cannot be understood. Since the ultra-wide band detecting signal has a very wide frequency bandwidth, and it is known that the propagation speed is different when the frequency changes. Therefore, it is not easy to tell which speed should be applied in the calculation. Theoretically, a specific speed value (corresponding to a frequency) must be selected as the speed of detecting signal which is based on analysis of frequency spectrum and this speed value must be able to reflect the character of the whole detecting signal.

(3) Targets are buried in an inhomogeneous surrounding. Both of the inhomogeneity of the earth for the ground-penetrating detection system, and the inhomogeneity of the tissues for the medical detection system can be thought as the inhomogeneity of electromagnetic parameters while the electromagnetic method is applied [10]. The propagation speed in medium can be written as:

$$v = \frac{1}{\sqrt{\frac{\mu\varepsilon}{2}\left[1 + \left(\frac{\sigma}{\omega\varepsilon}\right)^2 + 1\right]}} \qquad (2)$$

where $\varepsilon$, $\mu$, $\sigma$ represent the permittivity, magnetic conductivity, and electric conductivity, respectively. Unpredictability of these parameters of the medium will result in the imprecision of evaluation of the signal's propagation speed. In addition, reflections and refractions exit as the microwave signal travels in the inhomogeneous medium. In other words, the travelling path is unknown, thus the path length cannot be precisely estimated. As a result, neither the speed nor the distance can be accurately known.

The electromagnetic parameters of the medium can be set in our FDTD program, and so we can study the relation of inhomogeneity of the medium and contrast directly, which is given in table 1. With this table we can also find the equivalent translation for the inhomogeneity of medium and the TDPE. For instance, as the inhomogeneity of medium is 10% ($\varepsilon_r$=8.1~9.9, $\sigma$= 3.6 ~ 4.4 , for the typical value for breast tissues is $\varepsilon_r$ = 9 and $\sigma$= 0.4 ), it will result in a 9% TDPE and will give an imaging result with the contrast is 9 dB.

Table 1. Numerical relation of image contrast, inhomogeneity of the medium and TDPE

| Contrast ($S/\bar{N}$) dB | 11.86 | 10.4 | 9 | 7 | 5.8 | 5.15 | 4.37 |
|---|---|---|---|---|---|---|---|
| Inhomogeneity of medium % | 0 | 5 | 10 | 15 | 20 | 22 | 25 |
| TDPE % | 0 | 6 | 9 | 12 | 13 | 18 | 25 |

## Conclusion

The confocal microwave imaging algorithm requires to carry out the flight time of the backscattered signals from the target (or imaging) point to the detectors. We discussed three factors that may cause the image contrast decrease. It is pointed out in this paper that the inaccurate calculation of time-delay is the result of those three factors and is the final reason for the decrease of contrast. Finally, the numerical relation regarding the contrast and time delay percentage error is given. According to method we applied in this paper, the relationship of the time-delay error and the resolution, which is another very important parameter in imaging work, can also be studied.

## Afterword

The work presented in this conference paper was afterwards published on a journal [11] with more detailed analysis. Although lately there have been many advanced algorithms and prototypes

[12] developed to reconstructed images using collected microwave signals, the CMI approach is a simple but effective, also rapid method for the MMI study. Those more advance approaches increase the complexity in signal processing, often leading to longer computational time. In recent years, machine-learning algorithms have been applied to medical image reconstruction [13], assisted by more efficient simulation technique [14] to provide training data. Theses algorithms exhibit superpower such as super resolution, low noise, and fast speed, in solving inverse problems, surpassing the existing conventional algorithms. We are looking forward the machine-learning algorithms being applied to microwave breast imaging one day and give us surprises, and this will probably be real very soon!